\documentclass[aps,prl,twocolumn,amsmath,amssymb,amsfonts,10pt]{revtex4-2}

\usepackage{graphicx}
\usepackage{braket}
\usepackage[dvipsnames]{xcolor}
\usepackage[normalem]{ulem}
\usepackage[bookmarks=false,colorlinks=true,urlcolor=RoyalBlue,citecolor=NavyBlue,linkcolor=OrangeRed]{hyperref}
\usepackage{bm,bbm}

\begin{document}

\title{Local Markov Order and Global Inference in Many-Body Dynamics}
%\title{Learning and Memory in Many-Body Dynamics}

\author{Thomas Iadecola}
\email{iadecola@psu.edu}
\affiliation{Department of Physics, The Pennsylvania State University, University Park, Pennsylvania 16802, USA}
\affiliation{Institute for Computational and Data Sciences and Center for Theory of Emergent Quantum Matter, The Pennsylvania State University, University Park, Pennsylvania 16802, USA}
\affiliation{Materials Research Institute, The Pennsylvania State University, University Park, Pennsylvania 16802, USA}

\date{\today}

\begin{abstract}
We consider how the presence of conserved charges affects memory in a classical stochastic process, the symmetric exclusion process, with an observer constantly measuring a single site.
We find that the observer's measurement record becomes Markovian (i.e., loses memory) on a timescale that depends on their knowledge of the global charge, namely the total particle number. 
In particular, when the global charge is unknown \textit{a priori}, the observer's time series Markovianizes on a timescale constrained by their ability to learn it from their measurement record. 
Augmenting the observer's record with bulk measurements drives a charge-learnability transition between charge-fuzzy and -sharp phases. 
We show that the memory timescale tracks the learnability timescale, diverging in the fuzzy phase and remaining finite in the sharp phase.
\end{abstract}

\maketitle

The microscopic evolution of closed systems is local in time.
For an observer viewing only part of the system, however, the dynamics generically appears non-Markovian~\cite{Mori65,Zwanzig01,Grabert06}. 
Systems with conserved charges are a common setting where memory effects are important, mediated by slow hydrodynamic modes leading to algebraically decaying correlations~\cite{Alder70,Michaels75,Franosch11}.
These effects are also present in quantum systems~\cite{Breuer02}, where they pose challenges to efficient simulations~\cite{Foligno23,Vilkoviskiy25,Keeling25}.

In this work, we take an information-theoretic viewpoint on manifestations of non-Markovianity in many-body dynamics.
Consider an observer given access to a local patch of a large many-body system.
How much information about the past do they need to reconstruct the probability of a particular local configuration? 
How does this change when the observer is also transmitted information from beyond their local neighborhood?
The interplay between information acquisition and dynamics has become a topic of interest in the quantum community~\cite{Li18,Li19,Vasseur19,Skinner19,Gullans20,Zabalo20,Li23b,Ippoliti24}, where the learnability of global information from local measurements governs the ability to perform quantum error correction~\cite{Dennis02}.
This has given rise to a new class of learnability phase transitions driven by local measurements~\cite{Ippoliti24,Lee23,Fan24,Lee25}.
A prominent example is the charge sharpening phase transition, where the global charge goes from unlearnable to learnable as a function of the density of local measurements~\cite{Agrawal22,Barratt22a,Agrawal24}.
While this transition was originally uncovered in charge-conserving quantum dynamics, it also exists classically and manifests in the probability distribution of global charges conditioned on the measurement record~\cite{Barratt22b,McCulloch26}.
This places charge sharpening in a larger class of phase transitions in Bayesian inference that is also emerging~\cite{Kim25,Putz26,Nahum25}.

\begin{figure}[t!]
    \centering
    \includegraphics[width=\columnwidth]{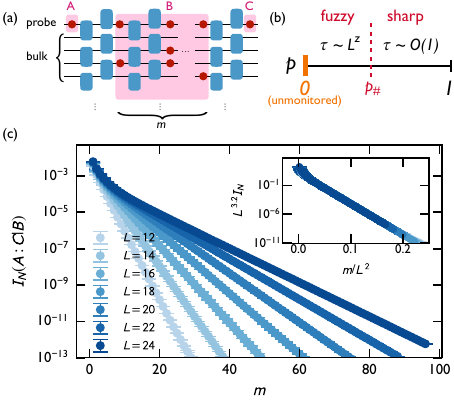}
    \caption{
    (a) Schematic of the model and measurement protocol. Blue boxes represent local SEP updates, and red circles represent measurements. Measurements in the highlighted regions $A$, $B$, and $C$ are used in computing the CMI.
    (b) Schematic phase diagram of the monitored SEP as a function of $p$, the measurement probability per time step. A charge sharpening (learning) transition between charge-fuzzy (unlearnable) and charge-sharp (learnable) phases occurs at $p=p_\sharp\approx 0.4$. We show that this transition manifests in a change in the scaling with system size $L$ of the Markovianization timescale $\tau$.
    (c) Decay of CMI in the unmonitored SEP as a function of temporal record length $m$ in a fixed charge sector $N=L/2$ for $L=12,...,24$. Inset: Finite-size scaling collapse following Eq.~\eqref{eq:I_N-scaling}.
    %for $z=2.00(1)$, $\beta=3.2(1)$ 2000 samples
    }
    \label{fig:fixed-p0}
\end{figure}

This work extends this line of inquiry into the time domain.
We consider a prototypical classical model of stochastic charge conserving dynamics, the symmetric exclusion process (SEP)~\cite{Spitzer70}, in which an observer monitors the charge on a single site and is optionally passed a record of measurements performed in the bulk of the system [see Fig.~\ref{fig:fixed-p0}(a) for a schematic].
The time-domain conditional mutual information (CMI) diagnoses the Markovianization timescale of the observer's record~\cite{Papapetrou13}.
We find that the decay of the CMI is constrained by hydrodynamic correlations and by the observer's ability to learn the global charge from the measurement record.
Below the charge sharpening transition, the Markovianization timescale diverges polynomially with system size, while above the transition it becomes order-one [see Fig.~\ref{fig:fixed-p0}(b)].
We expect that this connection between learning and memory is generic and can be probed in a variety of many-body systems using the approach developed here.
In fact, since the monitored SEP describes charge dynamics in the limit of large onsite Hilbert space in U(1)-symmetric random quantum circuits~\cite{Agrawal22,Barratt22a,Barratt22b}, our results apply directly in that setting as well.

\textit{Model and setup.}---The one-dimensional SEP describes $N$ hardcore particles hopping randomly on a chain of $L$ sites. 
The system's state is specified by the charge configuration $\bm n(t) = (n_0(t),\dots,n_{L-1}(t))$, with $n_j\in\{0,1\}$ and $|\bm n|=\sum_j n_j=N$. 
The configuration $\bm n(t+1)$ is obtained from $\bm n(t)$ by applying local stochastic updates: the local configurations $(n_j, n_{j+1})=(1,0)$ and $(0,1)$ are swapped or left invariant with equal probability, while $(0,0)$ and $(1,1)$ are frozen. 
These updates are applied to even and then odd $j$, with periodic boundary conditions. 
Equivalently, the SEP dynamics can be expressed in terms of the probability distribution $|q(t))=\sum_{\bm n} q_{\bm n}(t) |\bm n)$, with $q_{\bm n}(t)=(\bm n|q(t))$ the probability of configuration $\bm n$ after $t$ time steps. 
Viewed as a vector in the space of configurations, $|q(t))$, evolves under the transfer matrix
\begin{align}
    T = \!\!\prod_{j\text{ odd}}\!\!T_{j,j+1}\!\!\!\prod_{j\text{ even}}\!\!T_{j,j+1},\ \
    T_{j,j+1}=\begin{pmatrix} 1 & 0 &0 &0\\ 0 & \frac{1}{2} & \frac{1}{2} &0 \\ 0 & \frac{1}{2} & \frac{1}{2} &0\\ 0&0&0&1 \end{pmatrix}.
\end{align}
Conservation of probability is encoded in the inner product $(1|q(t)) = 1$, where $|1)=\sum_{\bm n}|\bm n)$ is the unnormalized uniform distribution over all allowed configurations (e.g., all configurations with $N$ particles).

The observer monitors the SEP dynamics by performing measurements (i.e., locally sampling from the underlying global distribution). 
We write the observer's measurement record as $r=\{r_{\bm j}(t)\}^{t_{\rm f}}_{t=0}$, where $r_{j_1}(t),\dots ,r_{j_k}(t)\in\{0,1\}$ are the results of measurements performed on sites $j_1,\dots,j_k$ at time $t$. 
The probability of observing measurement record $r$ is 
\begin{align}
\label{eq:prob}
    p(r) = (1|P_{r_{\bm j}(t_{\rm f})}T\dots P_{r_{\bm j}(1)}TP_{r_{\bm j}(0)}|q(0)),
\end{align}
where $P_{r_{\bm j}(t)}$ is a projection operator that zeroes out any entries in the probability vector that are not compatible with the measurement outcomes on sites $\bm j$ at time $t$.
We will consider the case where the observer monitors a single \textit{probe site} (say, $j=0$), obtaining a measurement record $r=\{r_0(0),\dots,r_0(m+1)\}$ [see Fig.~\ref{fig:fixed-p0}(a)].

To quantify whether the observer's measurement record is Markovian, we employ the CMI:
\begin{align}
\label{eq:cmi}
    I(A\!:\!C|B) = \sum_{b}p(b)\, I(A\!:\!C|B=b),
\end{align}
where $A,B,C$ are random variables and where
\begin{align}
\label{eq:cmi-sample}
    I(A\!:\!C|B=b)=\sum_{a,c}p(a,c|b)\ln\frac{p(a,c|b)}{p(a|b)p(c|b)}
\end{align}
is the mutual information between $A$ and $C$ given that a sample $b\sim B$ was observed.
In this context, we treat $r_0(0)$ and $r_0(m+1)$ as samples from $A$ and $C$, respectively, and condition on the intermediate record $b=\{r_0(1),\dots,r_0(m)\}$ [see Fig.~\ref{fig:fixed-p0}(a)].
$I(A\!:\!C|B)=0$ implies $p(c|a,b)=p(c|b)$, so $a$ provides no predictive information about the future measurement $c$ when $b$ is known.
For a classical stochastic process, this is equivalent to the statement that the process is Markovian of order $m$.
More generally, we will say that the measurement record has finite Markov order~\cite{Racca07,Papapetrou13,Taranto19a,Taranto19b} if
\begin{align}\label{eq:markov-order}
I(A\!:\!C|B)\sim e^{-m/\tau}
\end{align} 
with a memory timescale $\tau$ of order one.

\textit{Fixed charge.}---We first compute the CMI~\eqref{eq:cmi} starting from the stationary state at half filling, $|q(0)) = |1)_{N=L/2} \equiv \binom{L}{L/2}^{-1}\sum_{|\bm n| = L/2}|\bm n)$ [we will denote this CMI by $I_N(A\!:\!C|B)$, since it is calculated within a fixed charge sector].
Our numerical protocol samples intermediate records $b\sim B$ of length $m$ as follows.
Starting from an initial half-filled configuration $\bm n$ selected uniformly at random, we run a single SEP instance out to $t=m_{\rm max}+1$ timesteps.
We take the charge history of the probe site from times $1$ to $m$ as our sample of $b$.
We then numerically evaluate Eq.~\eqref{eq:prob} to obtain $p(r)=p(a,b,c)$ for each $a$ and $c$.
From this, we obtain the conditional probability $p(a,c|b)$ and the marginals $p(a|b)$ and $p(c|b)$ to evaluate Eq.~\eqref{eq:cmi-sample}. 
We repeat this for many samples of $b$ and estimate $I_N(A\!:\!C|B)$ from the sample average.

\begin{figure}[t!]
    \centering
    \includegraphics[width=\columnwidth]{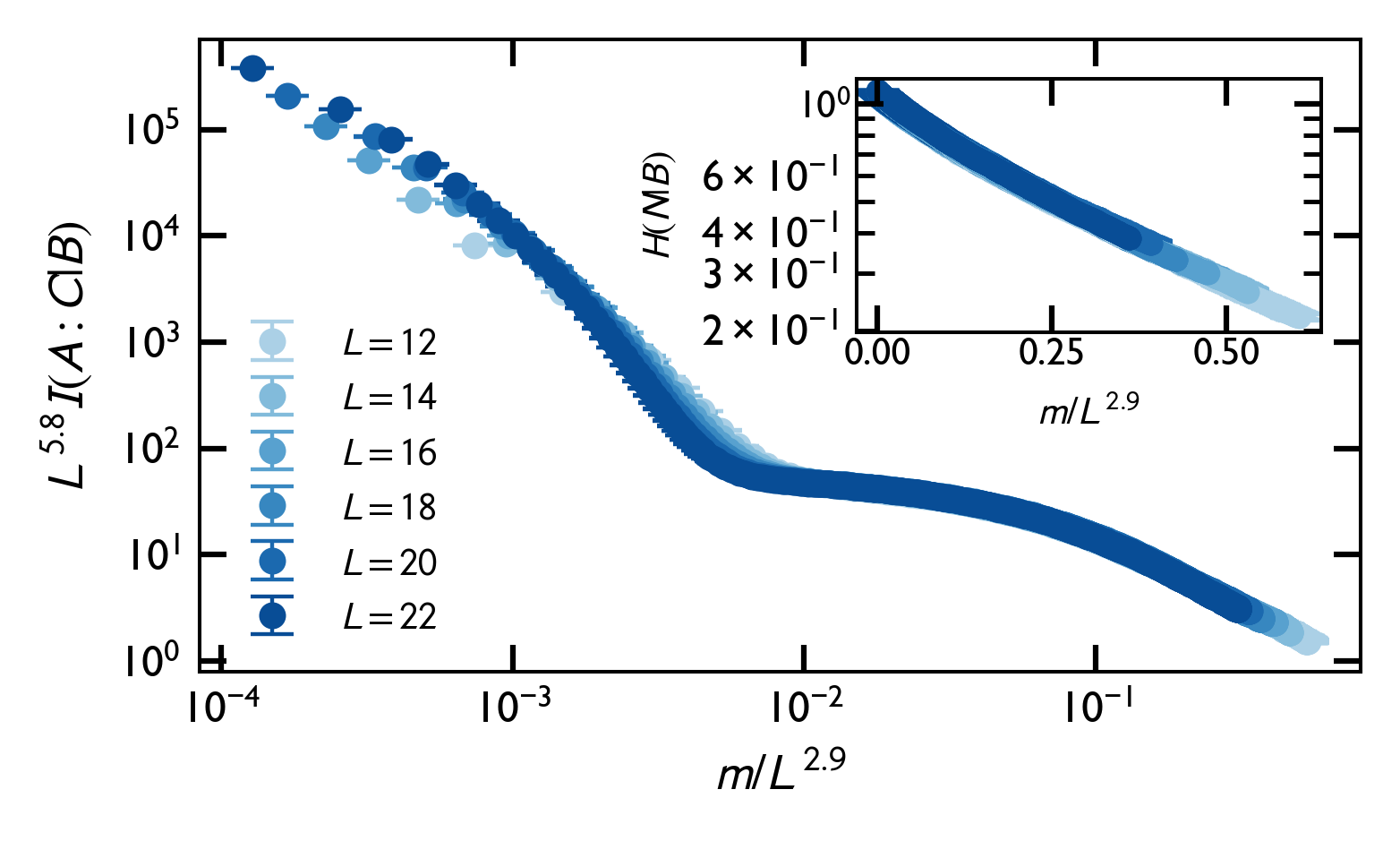}
    \caption{
    Decay of the CMI in the unmonitored SEP for an initial charge distribution spread over sectors $N=L/2$ and $L/2\pm 1$ with vertical and horizontal axes scaled to collapse the late-time tail. Inset: Decay of the posterior entropy \eqref{eq:posterior} with horizontal axis rescaled as in the main panel.
    %CMI $z=2.9(1)$, $\beta=5.8(1)$ entropy $z=2.90(5)$ 2000 samples
    }
    \label{fig:hidden-p0}
\end{figure}

\begin{figure*}[t!]
    \centering
    \includegraphics[width=0.9\textwidth]{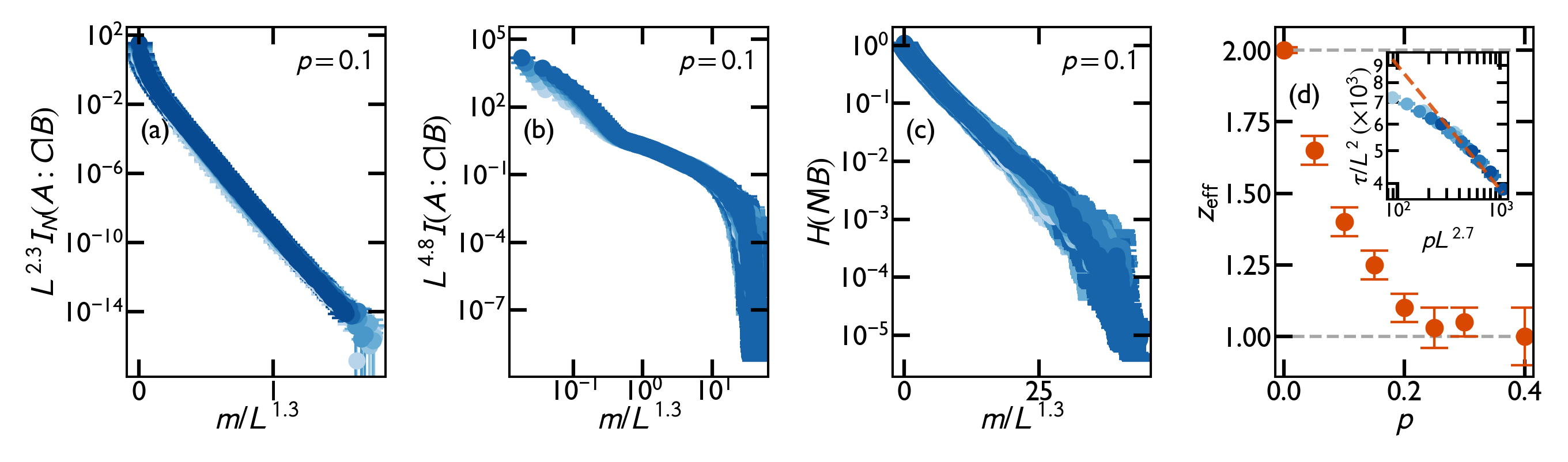}
    \caption{
    Markovianization and charge learning in the monitored SEP.
    (a) Decay of the CMI at half filling for bulk measurement rate $p=0.1$.
    (b) Decay of the hidden-charge CMI at $p=0.1$ with the same initial charge distribution as Fig.~\ref{fig:hidden-p0}.
    (c) Decay of the posterior entropy for the same initial charge distribution as in panel (b).
    (d) Flow of the effective dynamical exponent $z_{\rm eff}$ for the half-filling CMI from 2 to 1 as a function of $p \leq 0.4 \approx p_\sharp$.
    $z_{\rm eff}$ is extracted using the same $\chi^2$ analysis and sequences of system sizes as in Figs.~\ref{fig:fixed-p0} and \ref{fig:hidden-p0}.
    Inset: Finite-size flow of the exponential decay time scale $\tau(p,L)$ for system sizes $L=16,\dots,24$ and $p=0.0,0.05,\dots, 0.2$, collapsed according to Eq.~\eqref{eq:tau-scaling}. The orange dashed line is a guide for asymptotic scaling $\tau\sim L$.
    %(a) $z=1.35(5)$, $\beta=2.3(5)$ (b) $z=1.2(1)$, $\beta = 4.8(4)$ (c) $z=1.4(1)$ (d) $z=1.1(1)$ $\beta=2.5(5)$ (e) $z=1.2(2)$ $\beta=4.5(5)$ (f) $z=1.15(5)$
    }
    \label{fig:finite-p}
\end{figure*}

The results of these calculations, averaged over 2000 samples, are shown as a function of the temporal record length $m$ in Fig.~\ref{fig:fixed-p0}(c). 
We see a clear exponential decay in $m$ in line with Eq.~\eqref{eq:markov-order}, but with a decay timescale $\tau$ that increases with system size $L$. 
To extract the dependence on $L$, we perform a $\chi^2$ data collapse (inset) to a two-parameter scaling form
\begin{align}
\label{eq:I_N-scaling}
I_N(m,L) \sim L^\beta f_N(m/L^z),
\end{align}
which we assume to be valid for sufficiently long records $m\sim L^z$.
This collapse yields $z = 2.00(1)$, consistent with the diffusive charge transport in the SEP, with a scaling function $f_N(x)\sim e^{-x}$, and a finite-size scaling dimension $\beta = 3.2(1)$.
This indicates that the system has finite Markov order at any finite system size, but is intrinsically non-Markovian in the thermodynamic limit.
This is consistent with the expectation that slow hydrodynamic modes mediate long-lived memory, and indeed the memory timescale coincides with the natural hydrodynamic timescale.

We note in passing that the unconditioned mutual information (MI) between the time endpoints A and C also follows the finite-size scaling form~\eqref{eq:I_N-scaling} with $z=2$ scaling (see End Matter). The MI is expected to inherit a power-law scaling from the local charge autocorrelator $G(t)\sim 1/\sqrt{t}$ at late times. Establishing an analogous power-law behavior for the CMI, for example using field theory techniques, is an important topic for future work.

\textit{Hidden charge.}---The above results were obtained in a fixed charge sector, so the total charge $N$ is implicitly included in the observer's conditioning data.
On physical grounds, it is natural to ask what changes if the total charge is hidden from the observer.
Viewing the total charge as a hidden random variable, so that $I_{N=n}(A\!:\!C|B)\equiv I(A\!:\!C|B,N=n)$ one can show (see End Matter) that
\begin{align}\label{eq:cmi-charge-correlations}
    \!\!\!\!I(A\!:\!C|B)\!=\!I(A\!:\!C|B,N) \!+\! I(C\!:\!N|B) \!-\! I(C\!:\!N|A,B),
\end{align}
where $I(A\!:\!C|B,N)=\sum_{b,n}p(b,n)I_n(A\!:\!C|B=b)$, with $p(b,n)$ the joint distribution of the conditioning measurement record and the global charge.
In other words, the full CMI contains the fixed-charge contribution, considered above, and an additional piece governed by correlations between the fluctuating global charge and the final measurement outcome that have not been screened off by the measurement record.

To probe the full CMI with a fluctuating global charge, we initialize the system in the mixed-charge distribution $|q(0))=\sum_{N}w_N|1)_N$, where, as before, $|1)_N$ is the stationary charge distribution in charge sector $N$.
Here, we mix the charge over sectors $N=L/2$ and $L/2\pm 1$, introducing order-one charge fluctuations, with $w_N \propto \binom{L}{N}$ satisfying $\sum_N w_N=1$.
To sample an intermediate measurement record $b\sim B$, we draw a charge sector from the distribution $w_N$, and then draw a random configuration uniformly within that sector and evolve under a SEP instance.
We then compute $p_N(a,b,c)$ within each sector $N$ using Eq.~\eqref{eq:prob}, from which we obtain the fixed-charge conditional probabilities $p_N(a,c|b)$. 
The full conditional probability is then $p(a,c|b) = \sum_N p(N|b)\,  p_N(a,c|b)$, where
\begin{align}
p(N|b) = \frac{w_N \sum_{a,c}p_N(a,b,c)}{\sum_{N'} w_{N'} \sum_{a,c}p_{N'}(a,b,c)}
\end{align}
is the posterior charge distribution conditioned on record $b$.
The marginals $p(a|b)$ and $p(c|b)$ are then obtained and Eq.~\eqref{eq:cmi-sample} is evaluated as before.
Repeating this for many samples and taking the average yields the hidden-charge CMI $I(A\!:\!C|B)$.
We can also probe the observer's knowledge of the global charge by
tracking the Shannon entropy $H(x)=-\sum_x p(x)\ln p(x)$ of the posterior charge distribution $p(N|b)$:
\begin{align}\label{eq:posterior}
H(N|B) = \sum_b p(b) H(N|B=b).
\end{align}
This posterior entropy quantifies the observer's remaining uncertainty about the global charge given their measurement record.

In Fig.~\ref{fig:hidden-p0} we find that the hidden-charge CMI $I(A\!:\!C|B)$ (averaged over 2000 samples) decays with $m$ in a two stage process: an initial ``fast" transient and a slow long-time tail.
The transient regime exhibits a finite-size scaling collapse (not shown) in $m/L^2$, similar to the fixed-charge case, while the late-time tail collapses to a form like Eq.~\eqref{eq:I_N-scaling} with $\beta = 5.8(1)$ and $z= 2.9(1)$.
This parametrically longer decay timescale originates from a slow learning process in which the observer tries to reconstruct the global charge from its measurement record.
This learning process manifests in the decay of the posterior entropy $H(N|B)$ (inset), which exhibits a scaling collapse with $z=2.90(5)$.
The $z\approx 3$ scaling can be understood heuristically as follows: the observer needs to distinguish charge sectors $L/2$ and $L/2\pm 1$, which differ by a density of order $1/L$.
The background SEP dynamics is diffusive, so the integrated autocorrelation time $\tau_c\sim \int^{L^2}_0 dt\, G(t)\sim L$ sets the time needed for the observer to obtain a statistically independent sample of the density. 
A record of length $T$ therefore effectively contains $N_{s}\sim T/L$ samples, and $N_{s}\sim L^2$ such samples are needed to obtain a resolution $\sim 1/L$.
That the CMI decay is governed by this learning timescale $\sim L^3$ demonstrates that the unknown global charge provides the slowest-decaying memory channel in the observer's local measurement record.

% The late-time decays of the CMI and of the posterior entropy appear to be faster than power-law (see Fig.~\ref{fig:hidden-p0} main panel) but slower than exponential (see Fig.~\ref{fig:hidden-p0} inset).
% We find that the decay of the CMI is consistent with a stretched exponential $\sim e^{-a x^\alpha}$ for a broad range of stretching exponents $\alpha \approx 0.2$--$0.6$ as the data span less than two decades of the rescaled time $x=m/L^{2.9}$.
% Regardless, it is intriguing that the functional form of the decay is slower than the clearly exponential one observed for the fixed-charge CMI, and we leave a more precise determination of this functional form for future work.

\textit{Bulk monitoring.}---To further explore the connection between learning and memory, we now consider augmenting the observer's record with a finite density of measurement outcomes from the bulk of the system, away from the probe site.
Concretely, we now allow each site $j\neq 0$ to be measured with probability $p$ after each SEP step.
The measurement outcomes from a given monitored SEP trajectory are fed into Eq.~\eqref{eq:prob} as further conditioning data, so that $B$ now includes the observer's local measurement record and the spacetime locations and results of all bulk measurements.
In the absence of a continuously monitored probe site, this problem exhibits a learnability phase transition between a ``charge-fuzzy" phase at small $p$ and a ``charge-sharp" phase at large $p$~\cite{Agrawal22,Barratt22a,Barratt22b}.
In the fuzzy phase, the observer's record does not contain enough information to fully specify the global charge, while in the sharp phase it does.
We will see that the learnability timescale also constrains the Markovianization of the observer's measurement record.

In Fig.~\ref{fig:finite-p}, we show the fixed- (a) and hidden-charge CMI (b), along with the posterior entropy \eqref{eq:posterior} (c) for $p=0.1$, averaged over 2000 samples.
In panel (a), we see that the exponentially decaying form of the fixed-charge CMI observed at $p=0$ in Fig.~\ref{fig:fixed-p0} persists at finite $p$. 
Moreover, the two-stage decay of the hidden-charge CMI (b) also persists.
However, we find that both CMIs collapse with an effective dynamical exponent $z_{\rm eff}\approx 1.3$ that matches the one governing the decay of the posterior entropy.
That the fixed- and hidden-charge CMI collapse with consistent temporal finite-size scaling exponents is a consequence of the fact that bulk monitoring gives the local observer access to a finite density of charge measurement outcomes per time step.
This means the observer can estimate the density directly at each time step without waiting for the local measurement record to decorrelate.
Note that, although the decay timescales for all three quantities scale in the same way with system size, the decay timescale of the fixed-charge CMI remains parametrically shorter than that of the charge-indefinite observables.

\begin{figure}[t!]
    \centering
    \includegraphics[width=0.9\columnwidth]{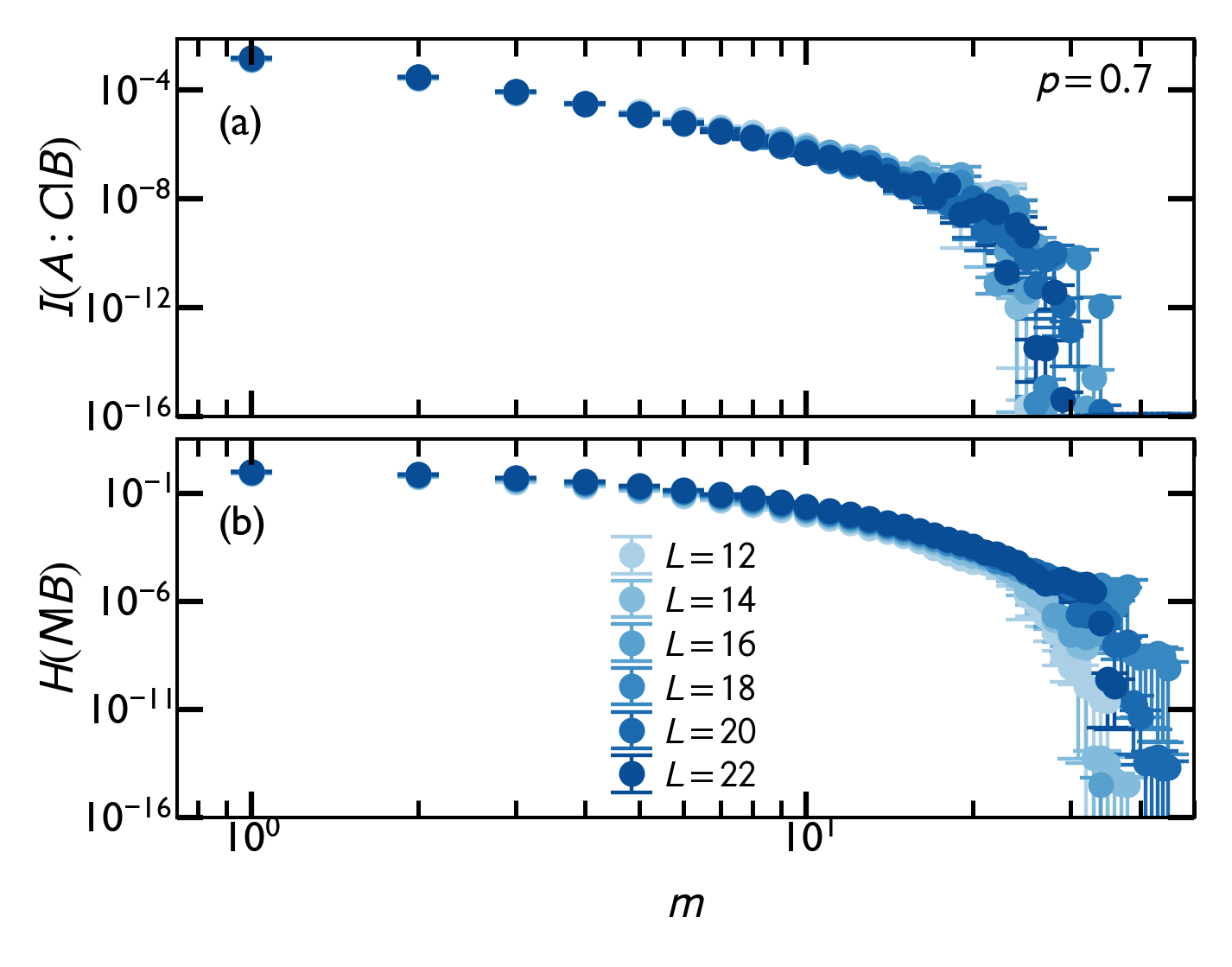}
    \caption{
    Decay of the hidden-charge CMI (a) and the posterior entropy (b) for $p=0.7$, with the same initial charge distribution as in Figs.~\ref{fig:hidden-p0} and \ref{fig:finite-p}.
    }
    \label{fig:large-p}
\end{figure}

In Fig.~\ref{fig:finite-p}(d), we show the evolution with $p$ of the effective dynamical exponent $z_{\rm eff}$ extracted from finite-size scaling collapse of the CMI at half filling, from the unmonitored limit $p=0$ towards the charge-sharpening transition at $p_{\sharp}\approx 0.4$~\footnote{Note that Ref.~\cite{Barratt22a} quotes $p_\sharp\approx 0.2$, but their monitoring protocol allows bulk measurements between each layer of gates, and therefore has approximately twice as many measurements per $p$ as our protocol.}.
We see that $z_{\rm eff}$ flows from 2 to 1 as $p$ increases towards $p_\sharp$.
We interpret this as a finite-size flow of the dynamical exponent from diffusive $z=2$ toward ballistic $z=1$ scaling, which is expected throughout the fuzzy phase at any finite $p$~\cite{Barratt22a}.
To corroborate this picture, we fit the fixed-charge CMI to the exponentially decaying form \eqref{eq:markov-order} for a range of $p$ and $L$ and collapse the extracted timescales $\tau(p,L)$ to the form
\begin{align}\label{eq:tau-scaling}
\tau(p,L)\sim L^2\, g(pL^\phi).
\end{align}
If $\tau\sim L$ as $L\to\infty$, the scaling function $g(x)\sim x^{-1/\phi}$ for large $x$ (orange dashed line).
The best-fit value $\phi = 2.7(1)$ produces a collapse consistent with this finite-size crossover form.
Deep in the charge-sharp phase, charge information is learned at a finite rate per unit volume and charged degrees of freedom become gapped~\cite{Barratt22a,Barratt22b}.
The expected $z=1$ scaling therefore gives way to an $O(1)$ timescale for charge learning and decay of correlations.
Our finite-size data are consistent with this expectation, as shown in Fig.~\ref{fig:large-p}, where the CMI and posterior entropy rapidly drop to zero within numerical precision on a timescale with no systematic $L$ dependence at the accessible system sizes.

\textit{Conclusion.}---We have studied the dynamics of local Markovianization and its relationship to hydrodynamics and information-theoretic transitions in the monitored SEP.
When the global charge is not knowable by the observer, the Markovianization timescale diverges with system size and tracks that of the slow charge learning process. 
In contrast, when the global charge is supplied as side information or learned at a finite rate, the slow hidden-charge memory channel is removed and the observer’s record can Markovianize on the shorter local relaxation timescale.
Our results highlight the connection between (monitored) hydrodynamics and local memory, and we expect that systems with conserved charges offer a universal setting where local memory persists in the thermodynamic limit.

One natural direction for future research is to generalize these considerations to quantum systems.
There, matters are complicated by the fact that a variety of local observables can be measured that may or may not reveal information about the local charges.
We expect charge-revealing measurements in purely unitary quantum dynamics to behave similarly to the U(1) case considered here.
More generally, Markovianization can be probed without reference to a choice of observable using the process tensor~\cite{Chiribella08,Chiribella09,Oreshkov12,Pollock18a,Pollock18b,Taranto19a,Taranto19b,Keeling25} or influence matrix~\cite{Lerose21,Foligno23,Vilkoviskiy25,Cerezo-Roquebrun26} formalism.
Finally, it is interesting to consider the relationship between temporal CMI and strong-to-weak spontaneous symmetry breaking (SWSSB)~\cite{Lee23,Lessa25} in open quantum systems, where diverging or algebraically decaying \textit{spatial} quantum CMI has been used as a diagnostic. 
We note that the charge-sharpening transition relevant to this work is an example of U(1) SWSSB~\cite{Huang25,Hauser26,Lee26,Tang26}.

\begin{acknowledgments}
\textit{Acknowledgments.}---I am grateful to Zhen Bi, Stefan Eccles, Sarang Gopalakrishnan, Xiantao Li, Jed Pixley, Sarah Shandera, Romain Vasseur, Ilya Vilkoviskiy, and Justin Wilson for helpful discussions.
I acknowledge the use of generative AI tools for coding and as support for data analysis, literature search, and manuscript revision.
This material is based upon work supported by the National Science Foundation under Grant No.~DMR-2611305.

% \textit{Data availability.}---The data that support the findings of this article are openly available~\cite{data}.
\end{acknowledgments}

\bibliography{refs}

\onecolumngrid
\bigskip
\bigskip
\begin{center}
\textbf{\large End Matter}
\end{center}
\vspace{0pt}

\twocolumngrid

\textit{Comparing MI and CMI at $p=0$.}---To contextualize our finite-size data for the CMI in the unmonitored SEP [Fig.~\ref{fig:fixed-p0}(c)], we study here the unconditioned MI between $A$ and $C$, $I(A\!:\!C)$, which is calculated from the joint probability distribution
\begin{align}
    p(a,c)=(1|P_{c}\,T^{m+1}P_{a}|q(0)).
\end{align}
In the late-time hydrodynamic regime, the decay of $I(A\!:\!C)$ is controlled by the charge autocorrelator $G(t)\sim 1/\sqrt{t}$.
More precisely, when $G(t)$ is small, we can write
\begin{align}
    I(A\!:\!C)\simeq \frac{G(t)^2}{2\bar{a}(1-\bar a)\bar c(1-\bar c)},
\end{align}
where $\bar a=p(A=1)$ and $\bar c = p(C=1)$ define the marginal distributions of the binary random variables $A$ and $C$.
This can be obtained by substituting the relation
\begin{align}
    p(a,c)&=p(a)p(c)+(-1)^{a+c}\, G(t)
\end{align}
into the definition of the mutual information,
\begin{align}
    I(A\!:\!C)=\sum^1_{a,c=0}p(a,c)\ln\frac{p(a,c)}{p(a)p(c)},
\end{align}
and expanding to leading order in $G(t)$.
Thus, if $G(t)\sim 1/\sqrt{m}$ for $t\sim m\gg 1$, we expect $I(A\!:\!C)\sim 1/m$.
Note that a similar treatment can also be applied to the CMI at fixed conditioning record, $I(A:C|B=b)$, to show it is governed by the square of the conditional correlation function $G(t|B=b)$.

\begin{figure}[t!]
    \centering
    \includegraphics[width=0.9\columnwidth]{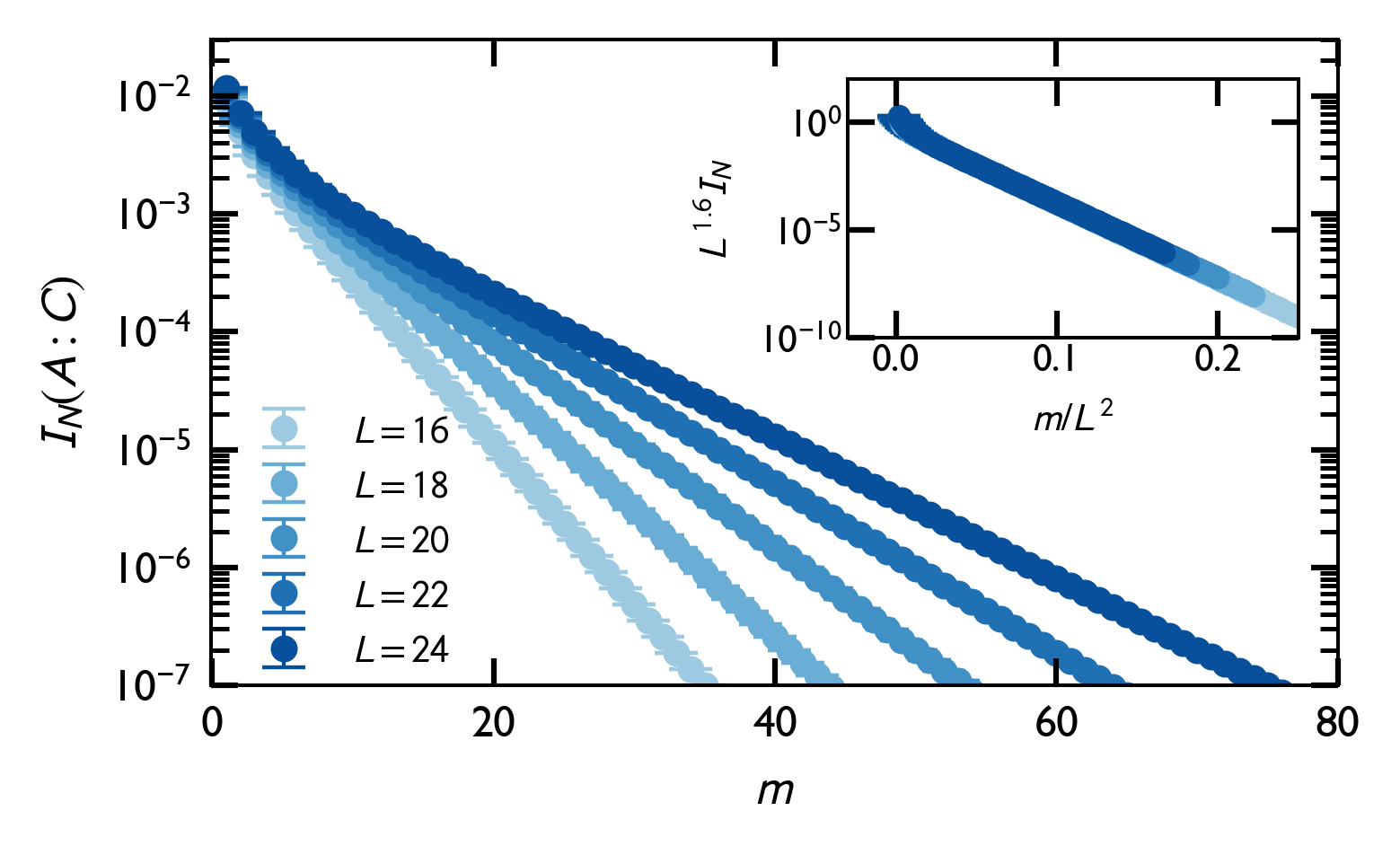}
    \caption{Decay of the unconditioned MI in the unmonitored SEP at half filling. Inset: Finite-size scaling collapse of the same data.
    }
    \label{fig:MI-fixed-p0}
\end{figure}

In Fig.~\ref{fig:MI-fixed-p0}, we plot the decay of the MI in the charge sector $N=L/2$ for the unmonitored SEP, in direct analogy to Fig.~\ref{fig:fixed-p0}. 
We see a clear exponential decay and a finite-size collapse (inset) with $z=2.03(5)$ and $\beta=1.6(1)$, consistent with diffusive hydrodynamic scaling. 
The significant difference in $\beta$ relative to $\beta_{\rm CMI}=3.2(1)$ may suggest that the MI and CMI probe different degrees of freedom in the long-wavelength limit---a question that could be clarified by looking at the MI and CMI in a field theory treatment.

\begin{figure}[b!]
    \centering
    \includegraphics[width=0.9\columnwidth]{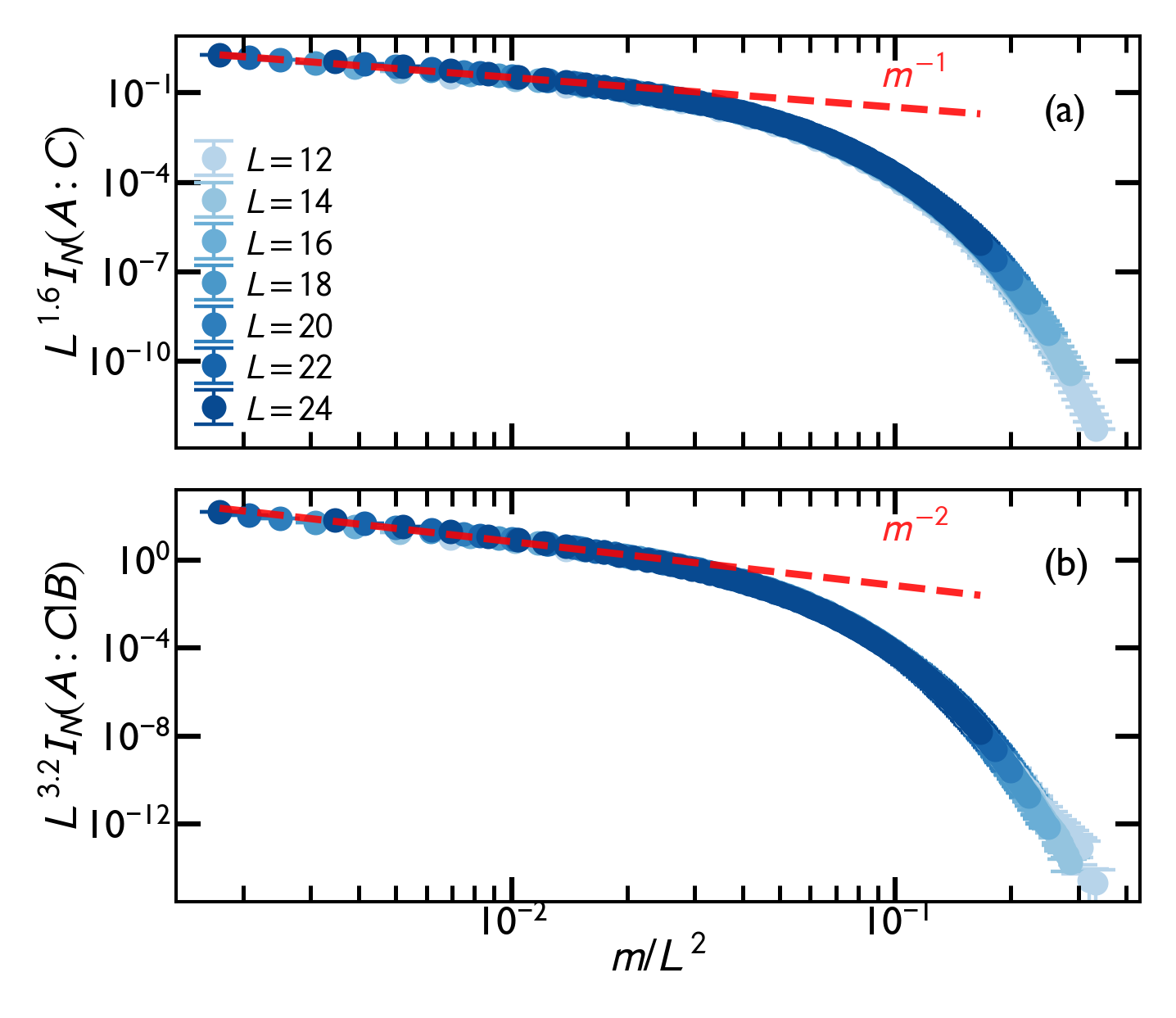}
    \caption{Collapsed data on logarithmic axes for (a) the MI and (b) the CMI in the unmonitored SEP at half filling (same data as Figs.~\ref{fig:MI-fixed-p0} and \ref{fig:fixed-p0}, respectively). Red dashed guide lines show the expected $1/m$ decay of the MI in (a), and a reference power-law decay $\sim 1/m^{2}$ of similar quality in (b).
    }
    \label{fig:CMI-MI}
\end{figure}

In Fig.~\ref{fig:CMI-MI} we plot (a) the MI and (b) the CMI in the charge sector $N=L/2$ for the unmonitored SEP, collapsed using the best-fit exponents.
The exponential decay manifests as substantial negative curvature on the logarithmic axes for sufficiently large $m\sim L^2$.
We expect hydrodynamic power-law decay for $m\ll L^2$, e.g. in the case where the thermodynamic limit $L\to\infty$ is taken from the outset.
Although our small size numerics do not allow access to a sufficiently large range of $m$ values to conclusively establish a power-law decay exponent in this regime, the small-$m$ behavior of the MI is consistent with the $\sim 1/m$ decay (red dashed guide line) expected based on the argument above.
The small-$m$ form of the CMI is similarly consistent with a power-law decay in the same range of $m$, albeit with a larger exponent $\sim 2$.
Further insight into the possible power-law scaling of the CMI in large systems could be obtained with approximate numerical methods (e.g. using matrix product state techniques) or by developing a field theory treatment of the CMI, both of which are interesting topics for future work.

\textit{Derivation of Eq.~\eqref{eq:cmi-charge-correlations}.}---The desired result follows from applying the chain rule for conditional mutual information to the CMI $I(C:A,N|B)$ two ways:
\begin{align}
    I(C:A,N|B) = I(C:N|B)+I(A:C|B,N)
\end{align}
and
\begin{align}
    I(C:A,N|B) = I(A:C|B)+I(C:N|A,B),
\end{align}
where we have used that $I(A:C|B)=I(C:A|B)$.
Equating these and solving for $I(A:C|B)$ gives Eq.~\eqref{eq:cmi-charge-correlations}.

\end{document}